%% file: dpf2009_JadrankaSekaric.tex
\documentclass[twocolumn,twoside,slac_two]{revtex4}
\usepackage{graphicx}
\usepackage{fancyhdr}
\usepackage{longtable}
\usepackage{multirow}
\def\Journal#1#2#3#4{{#1} {\bf #2}, #3 (#4)}
\def\AccJournal#1#2{{#1} (#2)}
\def\SubmJournal#1#2{{#1} (#2)}

\def\NPB{{\em Nucl. Phys.} B}
\def\PLB{{\em Phys. Lett.}  B}
\def\PRL{\em Phys. Rev. Lett.}
\def\PRD{{\em Phys. Rev.} D}

\def\met{{\mbox{$E\kern-0.57em\raise0.19ex\hbox{/}_{T}$}}}

\begin{document}

\title{Measurements of the Trilinear Gauge Boson Couplings from Diboson Production at D{\O}}

%

\author{J. Sekaric for the D{\O} Collaboration}
\affiliation{Department of Physics, Florida State University, Tallahassee, FL 32306-4350, USA}

\begin{abstract}
The most recent measurements of the trilinear gauge boson couplings from the diboson 
production at the D{\O} experiment has been presented.  The analyzed final states are 
$Z\gamma\rightarrow{\nu\nu}\gamma$, $WW\rightarrow{l}\nu{l'}\nu$, and 
$WW+WZ\rightarrow{l}\nu{jj}$.  We also present results obtained combining all 
final states involving the $W$ boson.  These results represent the most strigent limits 
set to date at the hadron collider.  
\end{abstract}

\maketitle

\thispagestyle{fancy}


\section{Introduction}
The simultaneous production of two vector bosons is a process of interest 
in many physics analysis at the Tevatron. Study of their production and 
interactions provide a test of the Electroweak Sector of the Standard Model 
(SM) either measuring thier production cross sections or trilinear gauge 
boson couplings (TGCs)~\cite{cit1}.  Any deviation from predicted SM values 
is an indication for New Physics (NP) beyond the SM and could give us some 
clues about the Electroweak Symmetry Breaking mechanism (EWSB).

\section{Phenomenology}
The TGCs contribute to diboson production via $s$-channel diagram.  Thus, 
production of $WW$ contains two trilinear $\gamma{WW}$ and $ZWW$ gauge boson 
vertices while the $WZ$ production contains the $ZWW$ vertex only.  The 
Effective Lagrangian wich describes $\gamma/ZWW$ vertices contains 14 
charged coupling parameters~\cite{cit2}. They are grouped according to 
the symmetry properties into $C$ (charge conjugation) and $P$ (parity) 
conserving ($g_{1}^{V},~{\kappa}_{V}$, and $\lambda_{V}$), $C$ and $P$ 
violating but $CP$ conserving ($g_{5}^{V}$), and $CP$ violating 
($g_{4}^{V},~\tilde{\kappa}_{V}$, and $\tilde{\lambda}_{V}$), 
where $V=Z,\gamma$.  In the SM all couplings vanish 
($g_{5}^{V}=g_{4}^{V}=\tilde{\kappa}_{V}=\tilde{\lambda}_{V}={\lambda}_{V}=0$)
except $g_{1}^{V}=\kappa_{V}=1$.  The value of $g_{1}^{\gamma}$ is
fixed by electromagnetic gauge invariance ($g_{1}^{\gamma}=1$) while
the value of $g_{1}^{Z}$ may differ from its SM value.  Considering
the $C$ and $P$ conserving couplings only, five couplings remain,
and their deviations from the SM values are denoted as the anomalous
TGCs $\Delta{g_{1}^{Z}}=(g_{1}^{Z}-1)$, $\Delta\kappa_{\gamma}=(\kappa_{\gamma}-1)$,
$\Delta\kappa_{Z}=(\kappa_{Z}-1)$, $\lambda_{\gamma}$ and $\lambda_{Z}$.  
Charged TGCs $g_{1}^{Z}$, $\kappa_{\gamma}$ and $\lambda_{\gamma}$ relate 
to the $W$ boson magnetic dipole moment $\mu_{W}$ and electromagnetic 
quadrupole moment $q_{W}$ as:
\begin{equation}
\begin{array}{l}
\mu_{W}= \frac{e}{2M_{W}}(g_{1}^{\gamma}+\kappa_{\gamma}+\lambda_{\gamma}), \\
q_{W}= -\frac{e}{M_{W}^{2}}(\kappa_{\gamma}-\lambda_{\gamma}).
\label{eqs1}
\end{array}
\end{equation}
In the $Z\gamma$ production, the $ZZ\gamma$ and $\gamma\gamma{Z}$ vertices 
contribute at the one-loop level or in the presence of NP but not at the 
tree-level.  The Effective Lagrangian contains 4 neutral coupling parameters 
$h_{i}^{V}$~\cite{cit3}, where $V=Z,\gamma$ and $i=1-4$.  The $h_{1}^{V}$ and 
$h_{2}^{V}$ violate $CP$ symmetry while $h_{3}^{V}$ and $h_{4}^{V}$ conserve 
$CP$ symmetry and all four couplings are equal to zero in the SM.  The couplings 
$h_{i}^{Z}$ also relate to the magnetic and electric dipole and quadrupole 
moments of the $Z$ boson~\cite{cit3}, but they were not experimentally studied 
at the D{\O} experiment yet.  

If anomalous TGCs are introduced in Effective Lagrangian, an unphysical 
increase in diboson production cross sections will result as the 
center-of-mass energy, $\sqrt{\hat{s}}$ approaches NP scale, $\Lambda_{NP}$.  
Such divergences would violate unitarity, but can be controlled by introducing 
a form factor $\Delta{a(\hat{s})}=\Delta{a_{0}}/(1+\hat{s}/\Lambda_{NP}^{2})^{n}$ 
for which the anomalous coupling vanishes as $\hat{s}\rightarrow\infty$.  The 
coupling $a_{0}$ is a low-energy approximation of the coupling $a(\hat{s})$, 
$n=2$ for $\gamma{WW}$ and $ZWW$ couplings, $n=3$ for $h_{1}^{V}$ and $h_{3}^{V}$, 
$n=4$ for $h_{2}^{V}$ and $h_{4}^{V}$.

\subsection{$Z\gamma\rightarrow{\nu\nu}\gamma$ Production}
\begin{figure}[h]
\centering
\includegraphics[width=80mm]{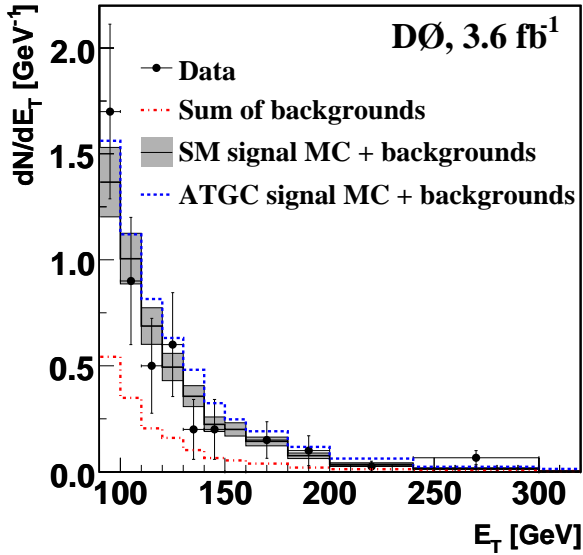}
\caption{Photon $E_{T}$ spectrum of $\nu\nu\gamma$ candidate events 
         compared to the SM signal and background, and the expected 
         distribution in the presence of anomalous TGCs. The systematic 
         and statistical uncertainties on the SM MC events are included 
         as shaded bands.} \label{fig1}
\end{figure}

The $Z\gamma\rightarrow{\nu\nu}\gamma$ events are reconstructed from 3.6 fb$^{-1}$ of D{\O} 
data.  Candidate events are required to have one isolated photon within pseudorapidity 
$|\eta_{det}|<$ 1.1~\cite{cit4}, transverse energy $E_{T}>$ 90~GeV and missing transverse 
energy $\met >$ 70~GeV.  The pointing algorithm~\cite{cit5} which provides the matching of 
the electromagnetic shower to the primary vertex is used in order to reduce the contribution 
from bremsstrahlung photons.  After all selection criteria were applied 51 $\nu\nu\gamma$ 
candidate events are observed.  The predicted numbers of signal and background events are 
$33.7\pm{3.4}$ and $17.3\pm{2.4}$, respectively.  The dominant background events are 
$W\rightarrow{e}\nu$ in which the electron is misidentified as a photon and it contributes 
with $9.7\pm 0.6$ events.  The measured cross section is 
$\sigma_{ZZ}\times{BR~(Z\rightarrow\nu\nu)}=32\pm{9}~(stat+syst)\pm{2}~(lumi)$ fb~\cite{cit6} which 
is in agreement with the next-to-leading (NLO) cross section of ($39\pm 4$) fb~\cite{cit7}.  The 
observed signal significance is 5.1 standard deviations (s.d.). For the TGC studies, the photon $E_{T}$ 
spectrum shown in~Figure~\ref{fig1} is used to set the limits on $ZZ\gamma$ and $Z\gamma\gamma$ 
couplings.  The Monte Carlo (MC) signal samples were generated with the leading-order (LO) $Z\gamma$ 
generator~\cite{LOzgamma}, corrected for the NLO effects with an $E_{T}-$dependent $K$ 
factor~\cite{cit7} and passed through a parameterized simulation of the D{\O} detector.  The signal 
tamplates with different anomalous TGC values are fitted to data in each bin of the photon $E_{T}$ 
distribution, together with the SM background.  The binned likelihood method~\cite{LLHzgamma} with 
the likelihood is used to extract the limits on TGCs from data.  

The one-dimensional 95$\%$ C.L. limits for ${h_{30,40}^{\gamma,Z}}$ at $\Lambda_{NP} = 1.5$~TeV are 
$|h_{30}^{\gamma}|<0.036$, $|h_{30}^{Z}|<0.035$ and $|h_{40}^{\gamma,Z}|<0.0019$~\cite{cit6}.  The 
combination with the previous 1 fb$^{-1}$ data analysis in $ll\gamma$ final states~\cite{RunIIAzgamma} 
results in the most restrictive limits on these couplings at 95$\%$ C.L. of $|h_{30}^{\gamma,Z}|<0.033$ 
and $|h_{40}^{\gamma,Z}|<0.0017$.  Three of them, $h_{40}^{\gamma},h_{40}^{Z}$ and $h_{30}^{Z}$, are 
world's best to date.

\subsection{$WW\rightarrow{l}\nu{l}\nu$ Production}
\begin{figure}[htb] 
\begin{centering}
\begin{tabular}{ccc}
\multirow{1}{*}[1.0in]{} & \includegraphics[width=70mm]{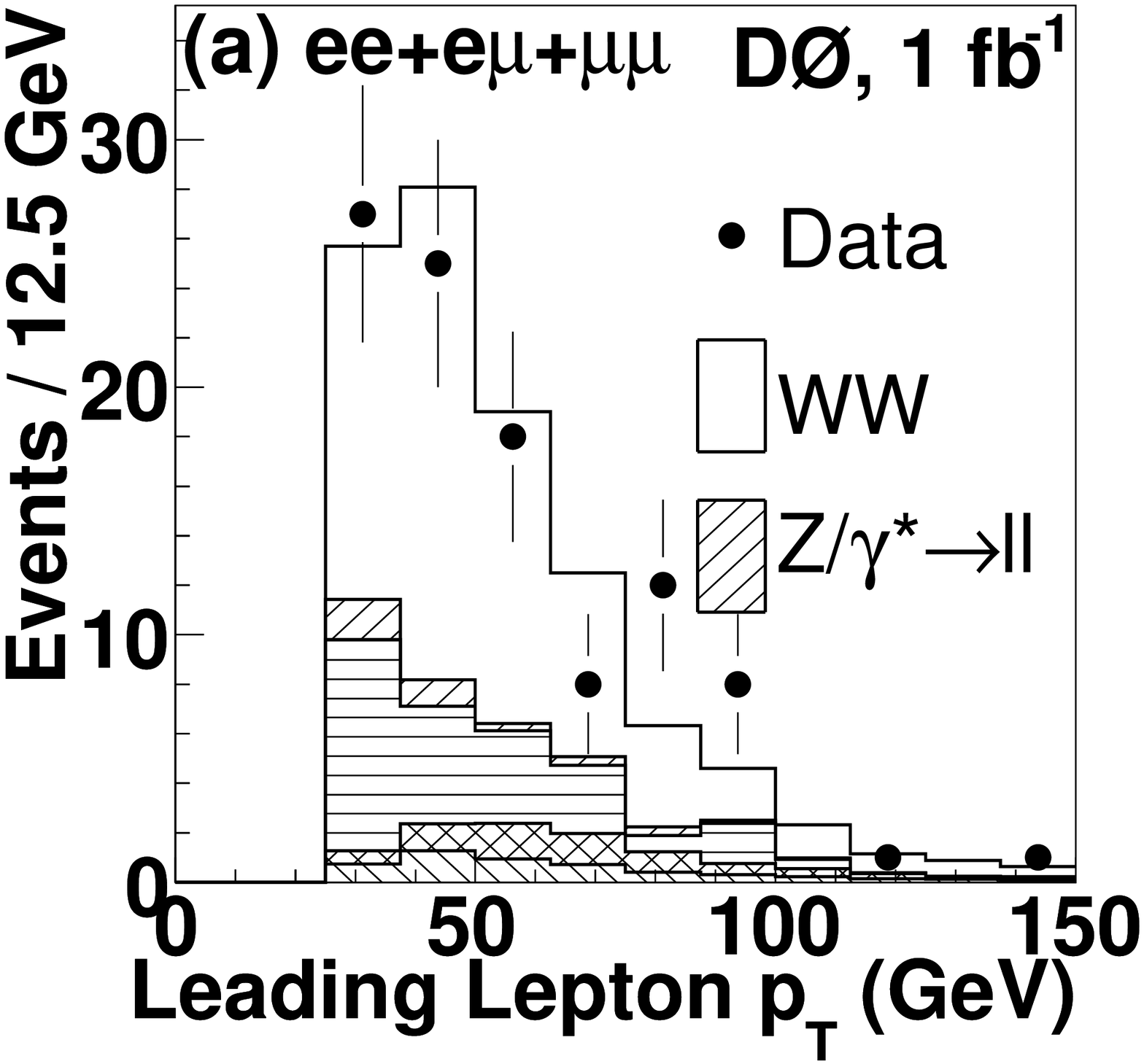}\\
\multirow{1}{*}[1.0in]{} & \includegraphics[width=70mm]{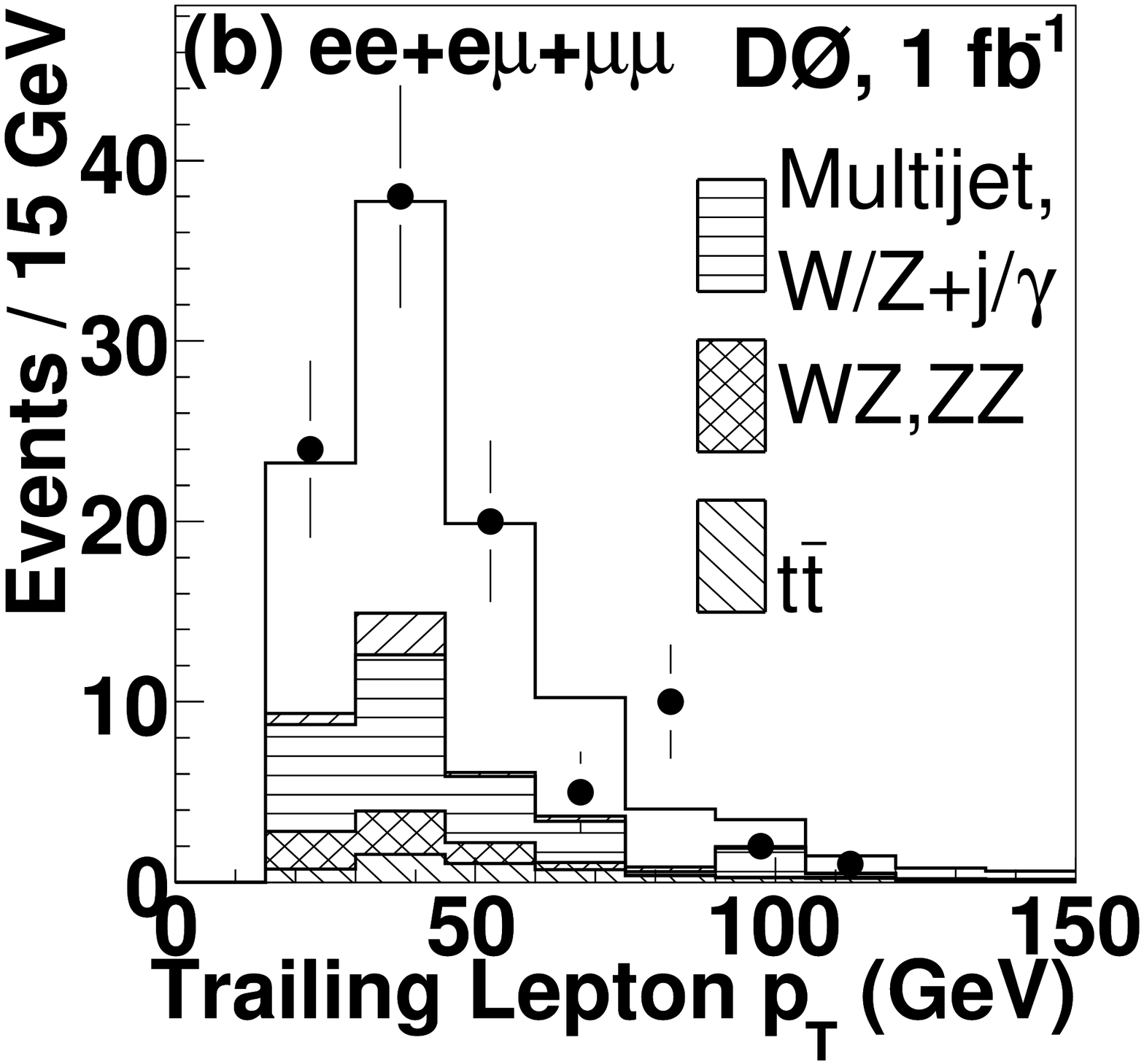}
\end{tabular}
\caption{Distributions of the (a) leading and (b) trailing lepton $p_{T}$ after final 
         selection in $WW\rightarrow{l}\nu{l'}\nu$ analysis, combined for all channels 
         ($ee+{e\mu}+{\mu\mu}$).  Data are compared to estimated signal, and background 
         sum.}
\label{fig2}
\end{centering}
\end{figure}

The most precise $WW$ cross section measurement at the D{\O} experiment is performed analyzing the 
$l\nu{l'}\nu$ ($l,l'=e,\mu$) final states with 1.0 fb$^{-1}$ of D{\O} data~\cite{cit8}.  In each $ll'$ 
final state ($ee,\mu\mu$ or $e\mu$) the two most energetic leptons are required to have $p_{T}>25~(15)$~GeV, 
to be of opposite charge and to be spatially separated from each other by $R>0.8$ ($ee$) and $R>0.5$ ($e\mu$).  
The $Z/\gamma^{*}\rightarrow{ll}$ background is effectively removed requiring $\met >$ 45~GeV ($ee$), 
20~GeV ($e\mu$) or 35~GeV ($\mu\mu$), $\met >$ 50~GeV if $|M_{Z}-m_{ee}|<$ 6~GeV ($ee$), 
$\Delta\phi_{\mu\mu}<2.45$ and $\met >$ 40~GeV if $\Delta\phi_{e\mu}>2.8$.  Imposing the upper cut 
on the transverse momentum of the $WW$ system, of 20~GeV ($ee$), 25~GeV ($e\mu$) and 16~GeV ($\mu\mu$) 
minimizes the $t\bar{t}$ background.  After all selection criteria were applied, all three combined channels 
yield 100 candidate events, $38.19\pm{4.01}$ predicted background events and $64.70\pm{1.12}$ predicted 
signal events.  The cross section measurements in the individual channels are combined, yielding 
$\sigma_{WW}=11.5\pm{2.1}~(stat+syst)\pm{0.7}~(lumi)$ pb which is in agreement with the SM NLO prediction of 
$12.4\pm{0.8}$ pb~\cite{cit9}.  The $p_{T}$ distributions of the leading and trailing leptons~(Figure~\ref{fig2}) 
were used to set limits on anomalous TGCs, considering two different parameterizations between the couplings.  
Requiring $SU(2)_L\times{U(1)_Y}$ gauge symmetry~\cite{cit10} gives the following relationship between the TGC 
parameters: $\Delta\kappa_{Z} = \Delta g^{Z}_{1}-\Delta\kappa_{\gamma}\cdot\tan^{2}\theta_{W}$ and 
$\lambda~\equiv~\lambda_{Z} = \lambda_{\gamma}$.  Requiring equality between the $WW\gamma$ and $WWZ$ vertices 
($WW\gamma=WWZ$)~\cite{cit1} such that $\Delta\kappa~\equiv~\Delta\kappa_{Z}=\Delta\kappa_{\gamma} $ and 
$\lambda~\equiv~\lambda_{Z}=\lambda_{\gamma}$ and $g_{1}^{Z}=1$ reduces the number of TGCs from three to 
two.  We use the LO MC generator by Hagiwara, Woodside, and Zeppenfeld (HWZ)~\cite{cit1,cit2} 
to simulate the changes in $WW$ production cross section and kinematics as TGCs are varied about their SM 
values.  At each point in TGC space, generated events are passed through a parameterized simulation of the D{\O} 
detector.  To increase the sensitivity to anomalous couplings, events are sorted by lepton $p_{T}$ into a 
two-dimensional histogram.  For each bin in lepton $p_{T}$ space, the change in the number of $WW$ events is 
parameterized by a quadratic function in $\Delta\kappa_{\gamma},~\lambda_{\gamma},~\Delta g^{Z}_{1}$ space 
or in $\Delta\kappa,~\lambda$ space, depending on the TGC relation under study.  In the first case, the 
third TGC parameter is fixed to its SM value.  The likelihood values from MC to data fit are fitted 
with a 6th order polynomial and the limits are determined by integrating the likelihood curve and/or surface. 
The one-dimensional 95\% C.L. limits for $\Lambda_{NP}=2$~TeV are $-0.54<\Delta\kappa_{\gamma}<0.83$, 
$-0.14<\lambda_{\gamma}=\lambda_{Z}<0.18$ and $-0.14<\Delta{g_{1}^{Z}}<0.30$ under the 
$SU(2)_{L}\times{U(1)_{Y}}$-conserving constraints, and 
$-0.12<\Delta\kappa_{\gamma}=\Delta\kappa_{Z}<0.35$ and $-0.14<\lambda_{\gamma}=\lambda_{Z}<0.18$ under 
the assumption that $\gamma{WW}$ and $ZWW$ couplings are equal.

\subsection{$WW+WZ\rightarrow{l}\nu{jj}$ Production}
\begin{figure}[htb]
\centering
\includegraphics[width=80mm]{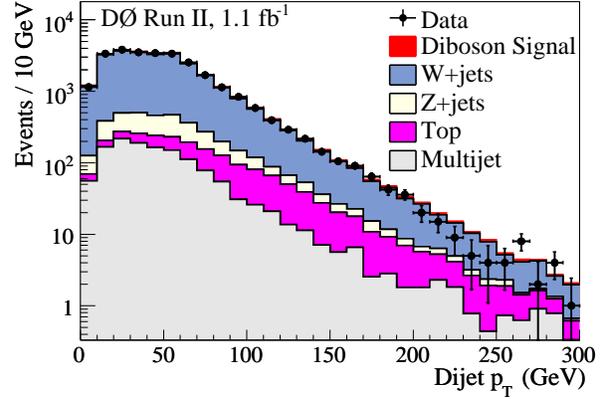}
\caption{The dijet $p_{T}$ distribution of combined (electron+muon) channels in the
         $WW+WZ\rightarrow{l}\nu{jj}$ analysis for data and SM predictions.} \label{fig3}
\end{figure}

Using 1.1 fb$^{-1}$ of D{\O} data the $l\nu{jj}$ ($l=e,\mu$) candidate events are selected requiring a 
single isolated lepton with $p_T>20$~GeV and $|\eta|<1.1\ (2.0)$ for electrons (muons), $\met >$ 20~GeV 
and at least two jets with $p_T>$ 20~GeV~\cite{cit11}.  The jet of highest $p_T$ must have $p_T>$ 30~GeV 
and the transverse mass of leptonically decaying $W$ boson must be $>$ 35~GeV to reduce the multijet 
background.  Because of the small signal-to-background ratio ($3\%$), an accurate modeling of the dominant 
$W$+jets background is essential and therefore, studied in great detail.  After all selection criteria were 
applied, the signal and the backgrounds are further separated using a multivariate classifier, Random Forest 
(RF)~\cite{cit12} for purposes of the cross section measurement.  The signal cross section is determined 
from a fit of signal and background RF templates to the data with respect to variations in the systematic 
uncertainties~\cite{cit13} and is measured to be $\sigma_{WW+WZ}=20.2\pm{2.5}~(stat)\pm{3.6}~(syst)\pm{1.2}~(lumi)$~pb 
which is consistent with the SM NLO prediction of $\sigma(WW+WZ)=16.1\pm{0.9}$~pb~\cite{cit9}.  The observed 
signal significance is 4.4 s.d..  The TGCs in $l\nu{jj}$ final states are measured from the dijet $p_{T}$ 
distribution in the combined electron and muon channels shown in~Figure~\ref{fig3}.  Data were compared 
to MC prediction with different anomalous TGC parameters. 
For this purpose we used the reweighting method to reweight the SM distribution to various anomalous TGC 
models~\cite{cit14} predicted by the HWZ generator using the weight $R$.  This method is based on the fact that 
the differential cross section has a quadratic dependence on the anomalous couplings and can be written as
\begin{equation}
\begin{array}{ccl} d\sigma & = & {\textit{const}}\cdot|\mathcal M|^{2}dX \\ 
& = & {\textit{const}}\cdot|\mathcal M|_{SM}^{2}\frac{|{\mathcal M}|^{2}}{|{\mathcal M}|_{SM}^{2}}dX \\ 
& = & {\textit{const}}\cdot|\mathcal M|^{2}_{SM}[1+A(X)\Delta\kappa+B(X)\Delta\kappa^{2}\\ 
& + & C(X)\lambda + D(X)\lambda^{2}+E(X)\Delta\kappa\lambda + ...]dX \\ 
& = & d\sigma_{SM}\cdot R(X;\Delta\kappa,\lambda,...)  
\label{expan} \end{array}{} \end{equation} 
\begin{table*}[htb]
 \begin{center}
   \begin{tabular}{l c c c}
   $SU(2)_L\times{U(1)_Y}$ scenario & $\Delta\kappa_{\gamma}$ & $\lambda=\lambda_{\gamma}=\lambda_{Z}$ & $\Delta{g_{1}^{Z}}$ \\ \hline\\
   $WZ\rightarrow \ell\nu \ell\ell$ ($1~\rm{fb^{-1}}$) & - & -0.17 $<\lambda <$ 0.21 & -0.14 $<\Delta{g_{1}^{Z}} <$ 0.34 \\
   $W\gamma\rightarrow \ell\nu\gamma$ ($0.7~\rm{fb^{-1}}$) & -0.51 $<\Delta\kappa_{\gamma} <$ 0.51 & -0.12 $<\lambda <$ 0.13 &  \\ 
   $WW\rightarrow \ell\nu \ell'\nu$ ($1~\rm{fb^{-1}}$) & -0.54 $<\Delta\kappa_{\gamma} <$ 0.83 & -0.14 $<\lambda <$ 0.18 & -0.14 $<\Delta{g_{1}^{Z}} <$ 0.30 \\ 
   $WW+WZ\rightarrow \ell \nu jj$ ($1.1~\rm{fb^{-1}}$) & -0.44 $<\Delta\kappa_{\gamma} <$ 0.55 & -0.10 $<\lambda <$ 0.11 & -0.12 $<\Delta{g_{1}^{Z}} <$ 0.20 \\ 
  \hline\\
  $WW\gamma=WWZ$ scenario & $\Delta\kappa_{\gamma}$ & $\lambda=\lambda_{\gamma}=\lambda_{Z}$ & $\Delta{g_{1}^{Z}}$ \\ \hline\\
  $WZ\rightarrow \ell\nu \ell\ell$ ($1~\rm{fb^{-1}}$) &  & -0.17 $<\lambda <$ 0.21 & \\ 
  $W\gamma\rightarrow \ell\nu\gamma$ ($0.7~\rm{fb^{-1}}$) &  & -0.12 $<\lambda <$ 0.13  & \\
  $WW\rightarrow \ell\nu \ell'\nu$ ($1~\rm{fb^{-1}}$) & -0.12 $<\Delta\kappa <$ 0.35 & -0.14 $<\lambda <$ 0.18 & \\ 
  $WW+WZ\rightarrow \ell\nu jj$ ($1.1~\rm{fb^{-1}}$) & -0.16 $<\Delta\kappa <$ 0.23 & -0.11 $<\lambda <$ 0.11&  \\ \hline
  \end{tabular}
  \end{center}
 \caption{Comparison of 95\% C.L. one-parameter TGC limits between the different channels 
          studied at D{\O} with $\approx{1}~\rm{fb^{-1}}$ of data: $WW\rightarrow \ell\nu \ell'\nu$,
          $W\gamma\rightarrow \ell\nu\gamma$, $WZ\rightarrow \ell\ell\ell\nu$ and
          $WW+WZ\rightarrow \ell\nu{jj}$ ($l=e,~\mu$) at $\Lambda_{NP}=$ 2~TeV.} \label{tab1}
\end{table*}
where $d\sigma$ is the differential cross section that includes the contribution from the anomalous couplings, 
$d\sigma_{SM}$ is the SM differential cross section, $|{\mathcal M}|^{2}$ is the matrix element squared in the 
presence of anomalous couplings, $|{\mathcal M}|_{SM}^{2}$ is the matrix element squared in the SM, $X$ is a 
kinematic distribution sensitive to the anomalous couplings and $A(X),~B(X),~C(X),~D(X)$, and $E(X)$ are 
reweighting coefficients dependent on $X$. In the $SU(2)_L\times{U(1)_Y}$ scenario $R$ is parametrized with 
the three couplings $\Delta{g_{1}^{Z}},~\Delta\kappa_{\gamma},~\lambda$ and nine reweighting coefficients, 
$A(X)-I(X)$ while in the $WW\gamma=WWZ$ scenario $R$ is parametrized with the two couplings 
$\Delta\kappa,~\lambda$ and five reweighting coefficients, $A(X)-E(X)$.  The kinematic variable $X$ is chosen 
to be the $p_{T}$ of the $q\bar{q}$ system, which is highly sensitive to anomalous TGCs.  To get signal MC 
templates with different anomalous TGCs, each event in a reconstructed dijet $p_{T}$ bin is weighted by the 
appropriate weight $R$ and all the weights are summed in that bin.  The observed limits are determined 
from a Poisson $\chi^{2}$ fit of background and reweighted signal MC distributions for different anomalous 
couplings contributions to the observed data with respect to variations to the systematic 
uncertainties~\cite{cit13} using the dijet $p_{T}$ distribution of candidate events.  For the 
$SU(2)_L\times{U(1)_Y}$ scenario, the most probable coupling values as measured in data with associated 
uncertainties at 68\% C.L. are $\kappa_{\gamma}=1.07^{+0.26}_{-0.29}$, $\lambda=0.00^{+0.06}_{-0.06}$, and 
$g_{1}^{Z}=1.04^{+0.09}_{-0.09}$.  For the $WW\gamma=WWZ$ scenario the most probable coupling values as measured 
in data with associated uncertainties at 68\%C.L. are $\kappa=1.04^{+0.11}_{-0.11}$ and 
$\lambda=0.00^{+0.06}_{-0.06}$.  The observed 95\% C.L. limits estimated from the single parameter fit are 
-0.44 $<\Delta\kappa_{\gamma} <$ 0.55, -0.10 $< \lambda <$ 0.11, and -0.12 $< \Delta g_{1}^{Z} <$ 0.20 for the 
$SU(2)_L\times{U(1)_Y}$ scenario or -0.16 $< \Delta\kappa <$ 0.23 and -0.11 $< \lambda <$ 0.11 for the 
$WW\gamma=WWZ$ scenario.  Table~\ref{tab1} shows the comparison of the 95\% C.L. limits on charged anomalous 
couplings set from data in different D{\O} diboson analyses.

This analysis yields the most stringent limits on $\gamma{WW}/ZWW$ anomalous couplings from the Tevatron to 
date, complementing similar measurements performed in fully leptonic decay modes from $W\gamma$, $WW$, and 
$WZ$ production.

\section{Combinaned Study of TGCs}
\begin{figure*}[htb]
\centering
\includegraphics[width=80mm]{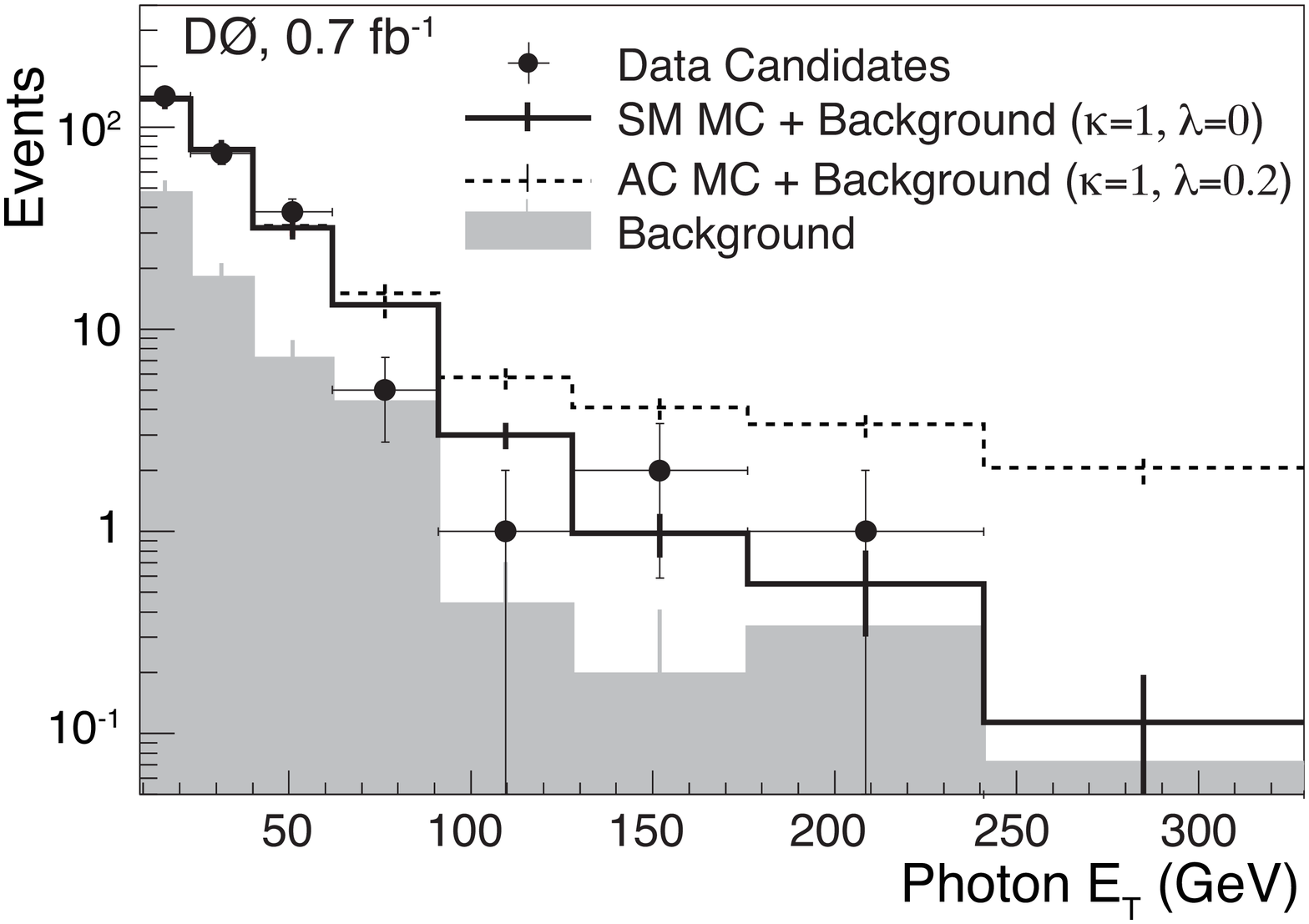}
\includegraphics[width=70mm]{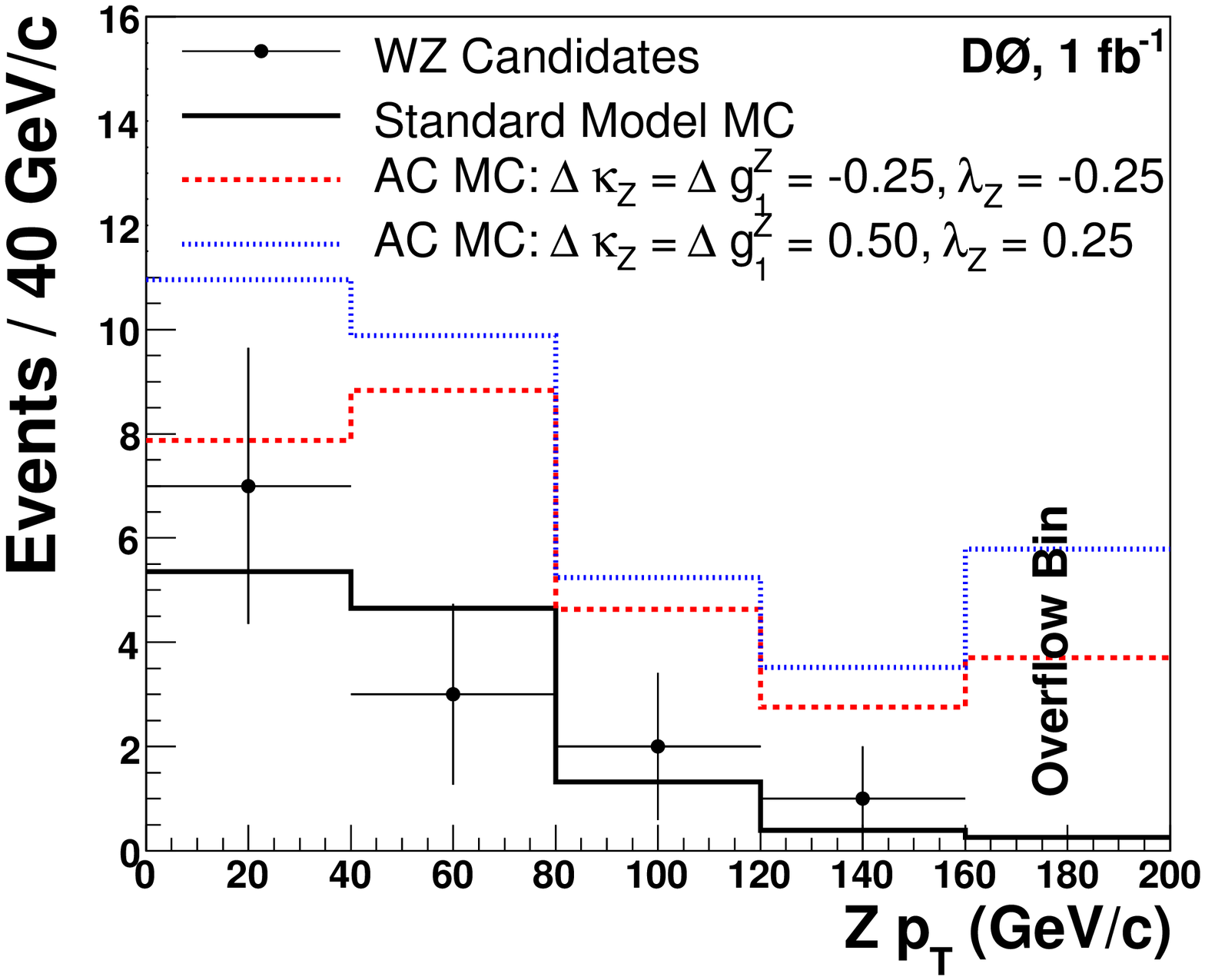}
\caption{Left: The photon $E_{T}$ spectra for the SM (solid line), an anomalous TGC point (dashed line), combined 
         electron and muon channel data candidates (black points), and the background estimate (shaded histogram)
         from the $W\gamma$ analysis. 
         Right: The reconstructed $Z$ boson $p_{T}$ of the $WZ$ candidate events.  The solid histogram is the 
         expected sum of signal and background in the SM and the black dots represent the data.  The dotted and 
         double dotted histograms are the expected sums of signal and background for two different cases of 
	 anomalous TGCs.} 
\label{fig4}
\end{figure*}

Finally, we combined four different diboson analyses, $WW\rightarrow{l}\nu{l'}\nu$, $WZ\rightarrow\nu{lll}$~\cite{cit15},
$W\gamma\rightarrow{l}\nu\gamma$~\cite{cit16} and $WW+WZ\rightarrow{l}\nu{jj}$, to set the limits on charged couplings 
using 0.7-1 fb$^{-1}$ of D{\O} data.  The kinematic distributions sensitive to anomalous TGCs are the photon $E_{T}$ 
spectrum in $W\gamma$ analysis and the $Z$ boson $p_{T}$ spectrum in $WZ$ analysis, shown in~Figure~\ref{fig4}.  The 
lepton $p_{T}$ distributions and dijet $p_{T}$ distribution are used as an input from $WW\rightarrow{l}\nu{l'}\nu$ and 
$WW+WZ\rightarrow{l}\nu{jj}$ analyses, respectively.  Two different relations between anomalous TGCs, 
$SU(2)_L\times{U(1)_Y}$ and $WW\gamma=WWZ$, has been considered for $\Lambda_{NP}=$ 2~TeV.  

Combined 95\% C.L. limits in both scenarios represent an improvement of about a factor of 3 relative to the 
previous D{\O}~\cite{cit17} and CDF~\cite{cit18} results and they are shown in~Table~\ref{tab2}.  These are the tightest 
limits to date on $\Delta\kappa_{\gamma},~\lambda_{\gamma}$ and $\Delta{g_{1}^{Z}}$ couplings at the hadron collider.  
The 68\% C.L. limits are presented in~Table~\ref{tab2} as well.  With only~1 fb$^{-1}$ of data the sensitivity obtained 
at the 68\% C.L. is comparable to that of an individual LEP2 experiment. 

We also measure the $W$ boson magnetic dipole and electric quadrupole moments respecting $SU(2)_L\otimes U(1)_Y$ symmetry 
with $g^Z_1=$ 1.  Their measured values and the one-dimensional 68\% C.L. intervals are $\mu_W=2.02^{+0.08}_{-0.09}~(e/2M_W)$ 
and $q_W=-1.00\pm{0.09}~(e/M^2_W)$, respectively.  This is the most stringent published result of $\mu_W$ and $q_W$ 
moments~\cite{cit19}. 

\begin{table}[htb]
\begin{tabular}{lccc}
\multicolumn{4}{c}{Limits for $SU(2)_L\otimes U(1)_Y$ scenario} \\
Parameter & Minimum & 68\% C.L. & 95\% C.L.\\
\hline
$\Delta\kappa_\gamma$	& $0.07$ & $[-0.13,0.23]$ & $[-0.29,0.38]$ \\
$\Delta g_1^Z$		& $0.05$ & $[-0.01,0.11]$ & $[-0.07,0.16]$ \\
$\lambda$	                & $0.00$ & $[-0.04,0.05]$ & $[-0.08,0.08]$ \\ \hline
\multicolumn{4}{c}{Limits for $WW\gamma=WWZ$ scenario} \\
Parameter & Minimum & 68\% C.L. & 95\% C.L.\\
\hline
$\Delta\kappa$	& $0.03$&$[ -0.04,0.11]$ & $[-0.11,0.18]$ \\
$\lambda$	& $0.00$&$[-0.05,0.05]$ & $[-0.08,0.08]$ \\ \hline
\end{tabular}
\caption{The most probable TGC values and one-dimensional 68\% and 95\% C.L. intervals on 
         anomalous values of $\gamma{WW}$ and $ZWW$ TGCs from the combined fit of
         $l\nu\gamma,~{l}\nu{l'}\nu$, $\nu{lll}$ and $l\nu{jj}$ final states.} 
\label{tab2}
\end{table}

\begin{figure*}[htb]
\centering
\includegraphics[width=80mm]{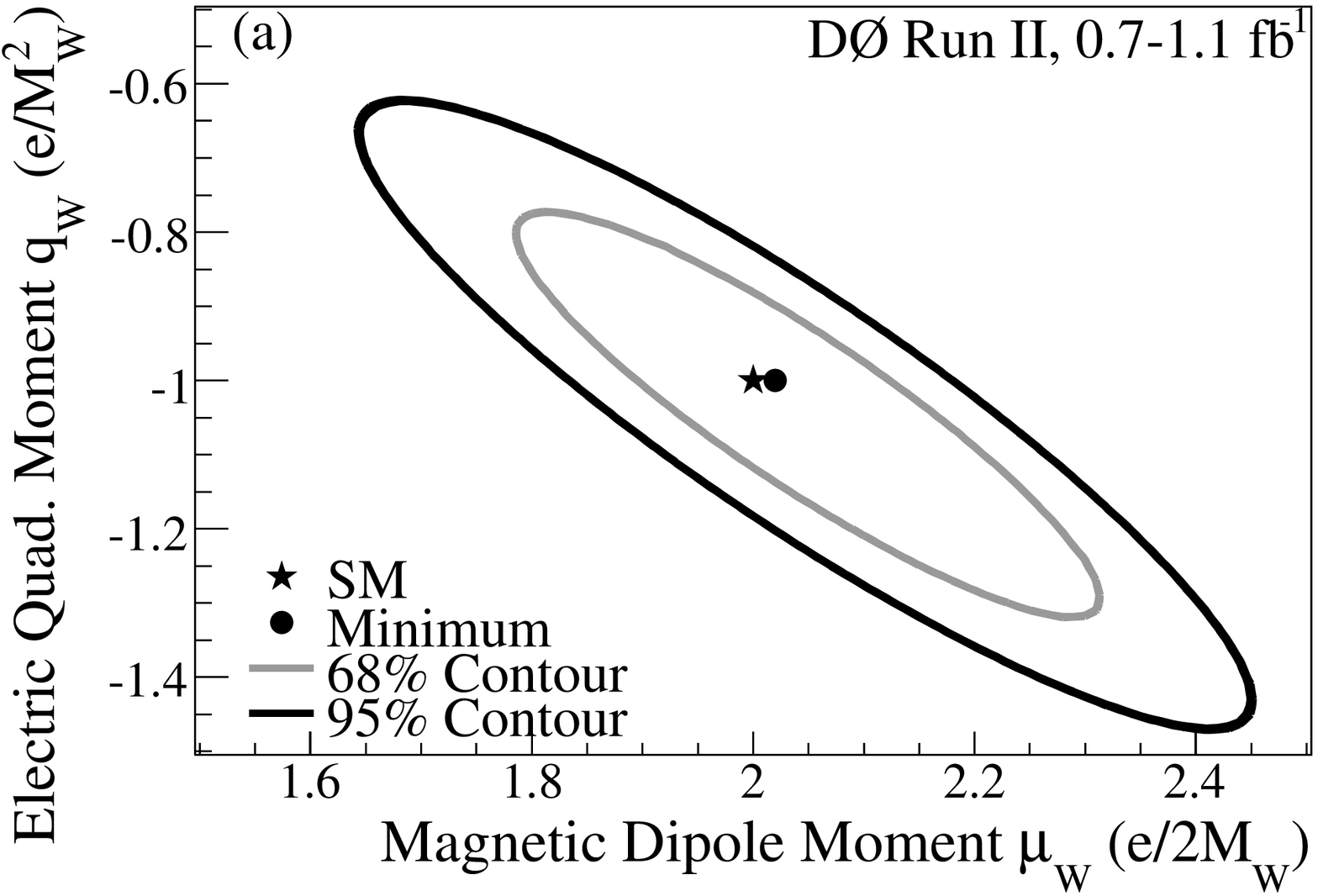}
\includegraphics[width=80mm]{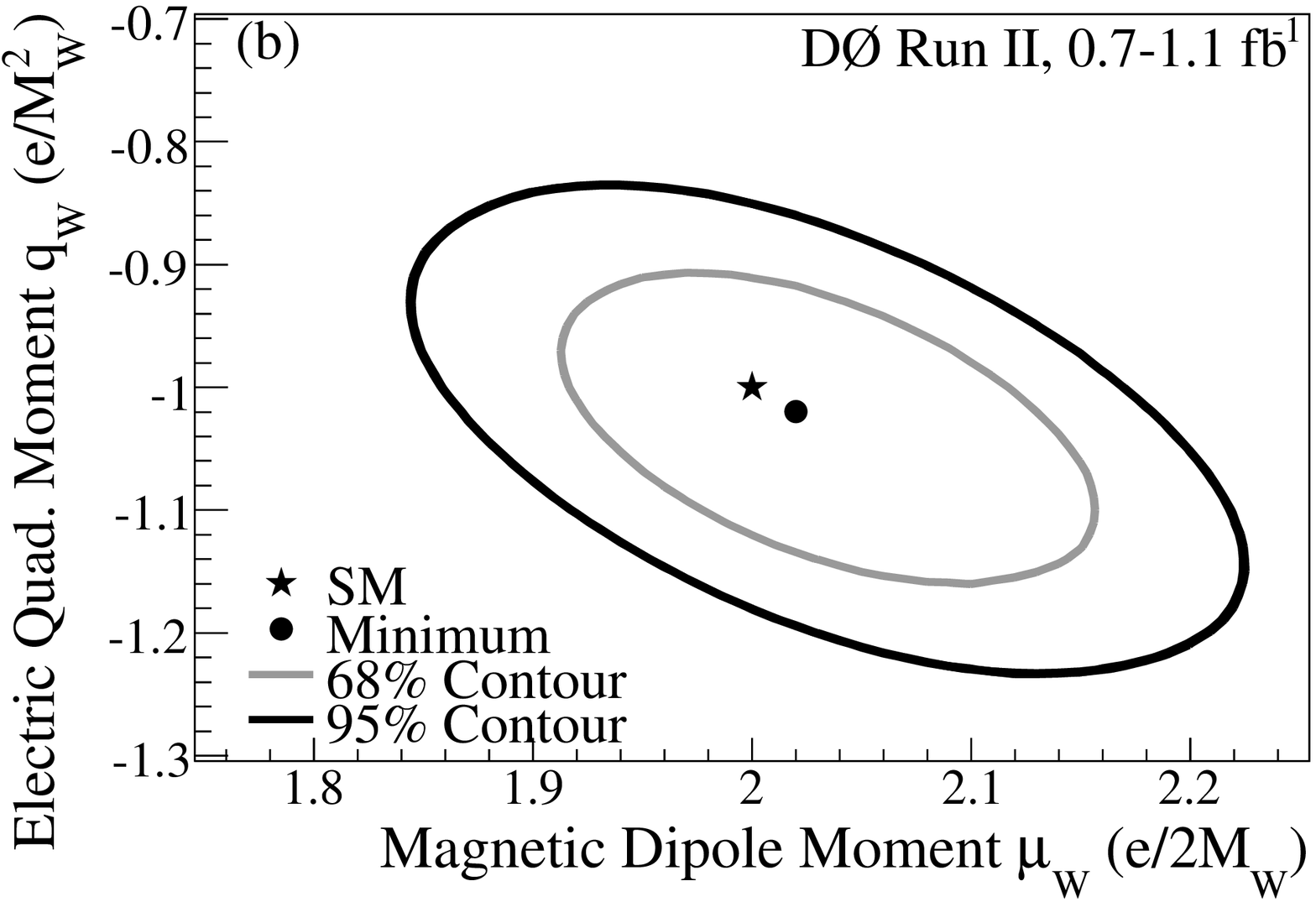}
\caption{Two-dimensional 68\% and 95\% C.L. limits for the $W$ boson electric 
         quadrupole moment vs. the magnetic dipole moment (a) when respecting 
         $SU(2)_L\times{U(1)_Y}$ symmetry and (b) when respecting $WW\gamma=WWZ$ 
         constraints. In both cases we assume $\Lambda_{NP}=$ 2~TeV.} 
\label{fig5}
\end{figure*}
\section{Conclusions}
The most resent measurements of the TGCs at the D{\O} experiment has been presented. No 
deviation from the SM prediction has been observed.  The D{\O} experiment sets the world's
tightest limits on neutral couplings $h_{30}^{Z,\gamma}$ and $h_{40}^{\gamma}$ in 
$Z\gamma\rightarrow\nu\nu\gamma$ analysis.  The limits set on $\Delta{g_{1}^{Z}},~
\Delta\kappa_{\gamma}$ and $\lambda_{\gamma}$ in the $l\nu{jj}$ final states are the 
stringent limits at the hadron collider obtained from only one final state.  The combination 
of $l\nu\gamma,~{l}\nu{l'}\nu$, $\nu{lll}$ and $l\nu{jj}$ final states, results in the 
tightest limits on charged TGCs at the hadron collider to date, significantly approaching 
to individual LEP2 sensitivities.  We also present the world's best results on the $W$ boson 
magnetic dipole and electromagnetic quadrupole moments.

\begin{acknowledgments}
\input acknowledgement_paragraph_r2.tex
\end{acknowledgments}

\bigskip 

\end{document}

%% file: acknowledgement_paragraph_r2.tex
%
We thank the staffs at Fermilab and collaborating institutions, 
and acknowledge support from the 
DOE and NSF (USA);
CEA and CNRS/IN2P3 (France);
FASI, Rosatom and RFBR (Russia);
CNPq, FAPERJ, FAPESP and FUNDUNESP (Brazil);
DAE and DST (India);
Colciencias (Colombia);
CONACyT (Mexico);
KRF and KOSEF (Korea);
CONICET and UBACyT (Argentina);
FOM (The Netherlands);
STFC and the Royal Society (United Kingdom);
MSMT and GACR (Czech Republic);
CRC Program, CFI, NSERC and WestGrid Project (Canada);
BMBF and DFG (Germany);
SFI (Ireland);
The Swedish Research Council (Sweden);
and
CAS and CNSF (China).

%% file: dpf2009_JadrankaSekaric.bbl
\begin{thebibliography}{99}
\bibitem{cit4} D{\O} uses a cylindrical coordinate system with the $z$ axis running along the  
               beam axis. Angles $\theta$ and $\phi$ are the polar and azimuthal angles, 
               respectively. Pseudorapidity is defined as $\eta = -\ln \left[ \tan(\theta/2) \right]$ 
               in which $\theta$ is measured with respect to the primary vertex. In the massless limit, 
               $\eta$ is equivalent to the rapidity $y = (1/2)\ln \left[ (E+p_z)/(E-p_z) \right]$. 
               $\eta_{\rm det}$ is the pseudorapidity measured with respect to the center of the detector.
\end{thebibliography}
